# All-electrical universal control of a double quantum dot qubit in silicon MOS

Patrick Harvey-Collard,[1,2] Ryan M. Jock,[2] N. Tobias Jacobson,[2] Andrew D. Baczewski,[2] Andrew M. Mounce,[2] Matthew J. Curry,[2,3] Daniel R. Ward,[2] John M. Anderson,[2] Ronald P. Manginell,[2] Joel R. Wendt,[2] Martin Rudolph,[2] Tammy Pluym,[2] Michael P. Lilly,[2] Michel Pioro-Ladrière[1] and Malcolm S. Carroll[2]

1. Département de physique et Institut quantique, Université de Sherbrooke, Sherbrooke, QC, J1K 2R1, Canada
2. Sandia National Laboratories, Albuquerque, NM, 87185, USA
3. Department of Physics and Astronomy, Center for Quantum Information and Control, University of New Mexico, Albuquerque, NM, 87131, USA

*Abstract*—Qubits based on transistor-like Si MOS nanodevices are promising for quantum computing. In this work, we demonstrate a double quantum dot spin qubit that is all-electrically controlled without the need for any external components, like micromagnets, that could complicate integration. Universal control of the qubit is achieved through spin-orbit-like and exchange interactions. Using single shot readout, we show both DC- and AC-control techniques. The fabrication technology used is completely compatible with CMOS.

## I. INTRODUCTION

Quantum computing promises to solve certain problems much faster than classical computing using quantum bits (qubits) [1]. Silicon metal-oxide-semiconductor (MOS) is a material system of choice for realizing qubits due to its compatibility with modern integrated circuit fabrication technology. Qubits in silicon are typically encoded in the spin degree of freedom of a single or a few electrons, each trapped in a quantum dot (QD). Isotopically enriched $^{28}$Si further provides a low-magnetic-noise environment to reduce qubit errors. Controlling single-electron spin-1/2 qubits requires the use of high frequency (> 20 GHz) signals and microwave antennae [2] or micromagnets [3], which can complicate integration. Multi-electron spin encodings can be operated with all-electrical, medium-frequency (~ 1 GHz) control signals at the expense of using up to three QDs per qubit [4]. An interesting compromise is the singlet-triplet (ST) two-electron encoding, which requires only two QDs and for which high-fidelity two-qubit gates are possible [5]. However, the ST qubit still requires an effective Zeeman energy difference $\Delta E_Z$ between the two QDs (equivalent to an effective magnetic field difference). In $^{28}$Si this can be provided by micromagnets [6] or donor atoms [7], but these schemes also suffer additional integration or maturity challenges.

In recent work, we have shown that although spin-orbit interaction is weak in bulk silicon, tight quantum confinement can lead to a surprisingly strong interface spin-orbit effect that is sufficiently large to be probed with a ST qubit [8]. In this work, we demonstrate universal qubit control and single shot readout of a ST qubit in silicon without any additional micromagnets or donor atoms. The effective $\Delta E_Z$ is consistent with a strong spin-orbit interaction resulting from the application of an external static magnetic field, which amounts to a difference in the electron g-factor between the two QDs. To demonstrate universal control, we first use traditional "DC-control" pulses, and then switch to a frame rotating at the qubit frequency where "AC-control" resonant pulses are used to implement arbitrary rotations around any axes. This technique simplifies the realization of complex pulse sequences.

## II. RESULTS

### A. Device and experiment

The device used for these experiments is shown in Fig. 1a. It consists of several poly-Si gate electrodes that are used to accumulate or deplete electrons at the Si-SiO$_2$ MOS interface (Fig. 1b). The fabrication technology used is completely compatible with CMOS, is metal-free and is based on plasma etching rather than lift-off processing. An epitaxial enriched $^{28}$Si layer was grown on top of the [001] wafer to provide a nuclear-spin-free environment for the spin qubit. Fabrication details are as in Ref. [8]. The device is cooled in a $^3$He/$^4$He dilution refrigerator with an electron temperature of about 300 mK. A magnetic field $B_{\text{ext}}$ = 1 T is applied in-plane with the device along the [100] crystallographic direction.

### B. Qubit readout

A single electron transistor (SET) is formed under the upper device gates that is used as a charge sensor (CS), see Fig. 1c. The CS is biased through a cryogenic SiGe heterojunction bipolar transistor (HBT) that is used as a current amplifier [9]. Single shot readout is achieved by using a differential current measurement to mitigate low-frequency noise and drift [4]. The detection contrast is improved using the "direct" enhanced latching readout described in Ref. [10]. During the readout, the two qubit states are distinguished by charge states that differ by one electron.

### C. Qubit control

A double QD is formed under the lower device gates. The QDs are located underneath the BS and BC gates. The number of electrons on each QD is denoted ($N_{\text{BS}}$, $N_{\text{BC}}$). The qubit is encoded in the singlet $|S\rangle = (|\uparrow\downarrow\rangle - |\downarrow\uparrow\rangle)/\sqrt{2}$ and unpolarized triplet $|T_0\rangle = (|\uparrow\downarrow\rangle + |\downarrow\uparrow\rangle)/\sqrt{2}$ of two electron spins [11], as shown in Fig. 2 and Fig. 3. The qubit is operated in the (1,1)

charge configuration, where the (1,1)$S$ and (1,1)$T_0$ qubit states are indistinguishable by the CS. The detuning $\varepsilon$, defined as the energy difference between the (2,0) and (1,1) states, is controlled by fast voltage pulses on gates BC and BL. When $\varepsilon$ is small, the residual wavefunction overlap between the (2,0) and (1,1) states generates "exchange" rotations of strength $J$ between the $|\uparrow\downarrow\rangle$ and $|\downarrow\uparrow\rangle$ states. At large $\varepsilon$, $J$ is suppressed and the qubit turns according to $\Delta E_Z$, which makes the two spins precess at different rates, effectively driving rotations between $|S\rangle$ and $|T_0\rangle$.

### D. Spin-orbit drive

In previous work, Jock *et al*. [8] have shown that despite spin-orbit interaction being weak in bulk silicon, surprisingly strong spin-orbit effects can arise from the quantum confinement against an interface in silicon QDs (Fig. 2). These effects can be understood in terms of an anisotropic electron g-tensor. The effective g-factor for each QD and for a given magnetic field depends on the microscopic details of the confinement and interface, yielding a $\Delta E_Z = (\Delta g)\mu_B B_{ext}$ that can be used to drive the ST qubit. Such an effect has been observed consistently in multiple devices.

In Fig. 4, we measure the qubit rotation frequency $f \sim \Delta E_Z/h$, where $h$ is the Planck constant, as a function of $B_{ext}$. We find a strong dependence on $B_{ext}$. However, the frequency is not directly proportional to $B_{ext}$ as in Ref. [8]. In this work, $B_{ext}$ is applied along the [100] crystallographic direction, for which spin-orbit effects are weaker. Independently, we have observed spin-orbit-induced transitions between the $|S\rangle$ and $|T_-\rangle$ states as in Nichol *et al*. [12]. The spin-orbit length extracted from these measurements is consistent with the work of Jock *et al*. [8], which supports a spin-orbit interpretation. Our work therefore suggests a residual spin-orbit effect that is not fully understood but can still be used for two-axis control.

### E. DC-control

The qubit is prepared in a (2,0)$S$ state by exchanging an electron with the reservoir. In Fig. 4, we demonstrate the ability to measure the qubit in two different bases using rapid adiabatic passage (RAP) or slow adiabatic passage (SAP) [11]. RAP corresponds to an "instantaneous" change of the detuning from the point of view of the spin degree of freedom, and thus allows measurement of the qubit in the $\{|S\rangle, |T_0\rangle\}$ basis. SAP corresponds to an adiabatic evolution of the spin Hamiltonian mapping $|S\rangle$ to $|\uparrow\downarrow\rangle$ and $|T_0\rangle$ to $|\downarrow\uparrow\rangle$, and thus allows a measurement of the qubit in the $\{|\uparrow\downarrow\rangle, |\downarrow\uparrow\rangle\}$ basis. In Fig. 5 and Fig. 6 we plot rotations around $J$ and $\Delta E_Z$, thus showing two-axis control of the qubit.

### F. AC-control

We now demonstrate an alternative "AC-control" scheme that simplifies the operation of the qubit [13, 14]. Such a scheme was introduced in GaAs-based ST qubits but has yet to be demonstrated in silicon. The scheme leverages the SAP and effectively encodes the qubit in the $\{|\uparrow\downarrow\rangle, |\downarrow\uparrow\rangle\}$ basis. Modulating the detuning at the qubit frequency modulates the $J$ term in the Hamiltonian, which amounts to an electron-spin-resonance-like control. The rotation axis is controlled with the phase of the pulse and the rotation angle by its duration. In the reference frame rotating at the qubit frequency, the qubit state is static between gates, which is convenient for complicated gate sequences. In Fig. 7, we plot the triplet return probability as a function of the frequency and pulse duration. A characteristic chevron pattern is observed centered on the resonance frequency.

## III. CONCLUSION

Our work demonstrates that all-electrical AC-control of spin qubits is possible in silicon MOS quantum dots without the integration of bulky components or non-standard CMOS processes. The driving mechanism is consistent with a spin-orbit effect that arises from the confinement at the interface. Optimizing the magnetic field orientation with the crystallographic axes could yield much faster qubit operation leading to reduced errors. This spin-orbit-driven ST qubit could dramatically simplify the integration of these devices and provide the building blocks of quantum hardware based on currently-available silicon MOS technology.

### ACKNOWLEDGEMENTS

This work was performed, in part, at the Center for Integrated Nanotechnologies, an Office of Science User Facility operated for the U.S. Department of Energy (DOE) Office of Science. Sandia National Laboratories is a multimission laboratory managed and operated by National Technology and Engineering Solutions of Sandia, LLC, a wholly owned subsidiary of Honeywell International, Inc., for the DOE's National Nuclear Security Administration under contract DE-NA0003525.

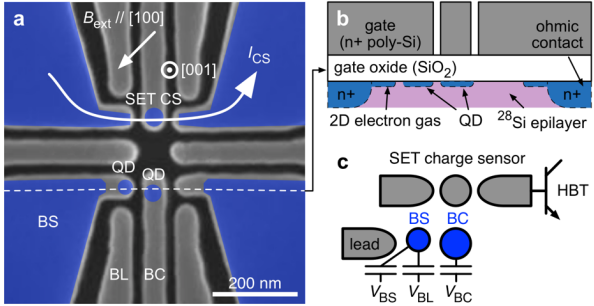

Figure 1: **(a)** Scanning electron microscope image of a device similar to the one measured showing the poly-Si gate structure. The blue overlay represents the regions where electrons accumulate below the Si-SiO$_2$ interface. The device is entirely based on CMOS technology and can thus be readily fabricated with existing foundry tools. **(b)** Schematic cross-section of the device stack (not to scale). The gate oxide is 35 nm thick, the poly-Si gates 200 nm. The active silicon region is enriched $^{28}$Si with 500 ppm residual $^{29}$Si. **(c)** The SET charge sensor (CS) is connected to a cryogenic HBT for fast charge sensing of the double QD occupancy. During qubit readout, the spin states are mapped to charge states that differ by one electron using spin blockade and an enhanced latching readout.

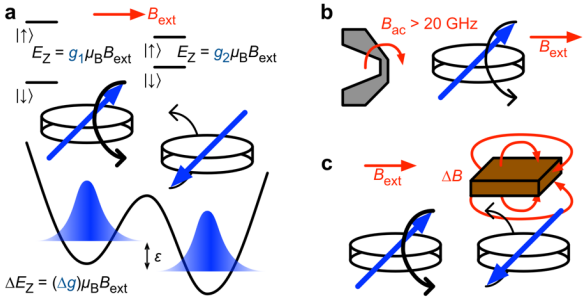

Figure 2: **(a)** The built-in spin-orbit interaction amounts to an effective electron g-factor that is different for the two QDs and makes the spins precess at different rates. The detuning $\varepsilon$ is defined as the energy difference between the (2,0) and (1,1) states. **(b-c)** Other spin qubits require antennae, micromagnets, or extra QDs to control the qubit. Our design features only two QDs and no other external components for all-electrical control.

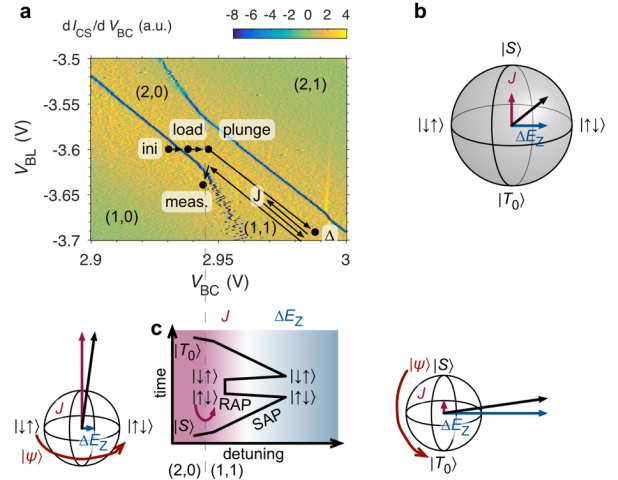

Figure 3: **(a)** Charge stability diagram showing the derivative of the CS current. The voltages on gates BC and BL control the charge occupation of the double QD system. Overlaid is a typical qubit preparation, control and readout pulse sequence. **(b)** Bloch sphere representation of the singlet-triplet (ST) spin qubit states. $|S\rangle$ and $|T_0\rangle$ can act as the 0 and 1 of the computational basis. The qubit state $|\psi\rangle$ evolves around the complex sum of $J$ and $\Delta E_Z$. **(c)** The detuning, controlled with fast voltage pulses, modulates the relative strength of $J$ and $\Delta E_Z$. Shown here is a typical exchange pulse sequence. The initial state $|S\rangle$ is mapped to $|\uparrow\downarrow\rangle$ through SAP at point $\Delta$. A RAP pulse to the J point triggers rotations between $|\uparrow\downarrow\rangle$ and $|\downarrow\uparrow\rangle$. The mapping is reversed through a SAP to the measurement point.

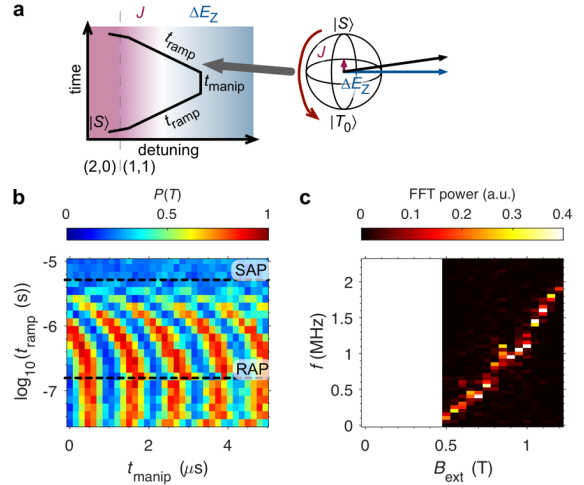

Figure 4: **(a)** Pulse sequence to determine the SAP and RAP ramp rates. **(b)** Triplet return probability $P(T)$ as a function of the manipulation time $t_{manip}$ and ramp time $t_{ramp}$. For short $t_{ramp}$ a RAP is achieved, which generates rotations between $|S\rangle$ and $|T_0\rangle$. For long $t_{ramp}$ a SAP maps $|S\rangle$ to $|\uparrow\downarrow\rangle$ and back, which in principle results in a trivial operation. **(c)** Fast Fourier transform (FFT) power of $P(T)$ data showing the frequency $f$ of the $\Delta E_Z$ rotations versus external magnetic field $B_{ext}$.

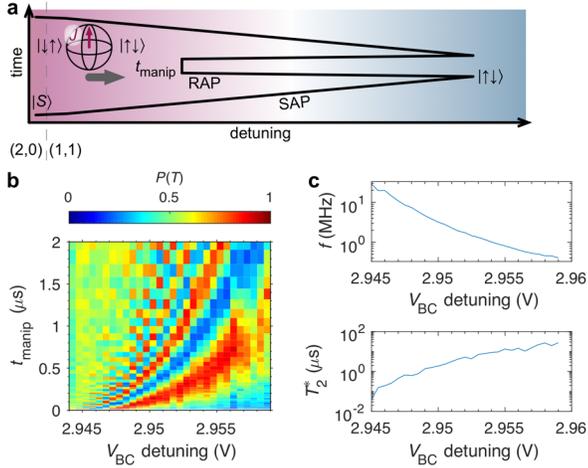

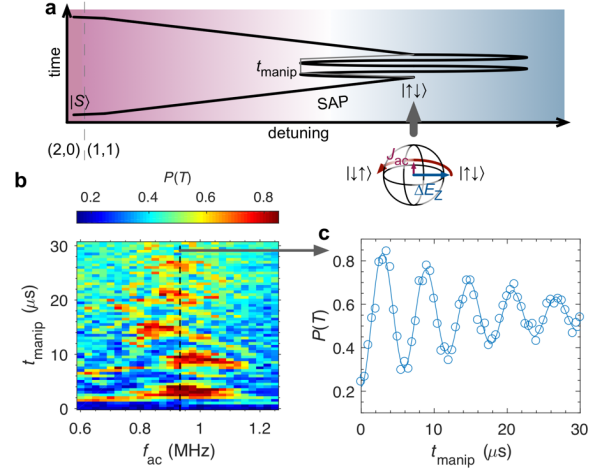

Figure 5: **(a)** Pulse sequence for DC-control of $J$ rotations in (b). **(b)** Coherent exchange rotations in $P(T)$ versus the $V_{BC}$ coordinate of the detuning pulse and the manipulation time $t_{manip}$. **(c)** Frequency $f \sim J/h$ and coherence time $T_2^*$ of the rotations extracted from fits to the data in (b) of the form $P(T) = \frac{v}{2}\exp\left(-(t_{manip}/T_2^*)^2\right)\sin(2\pi f + \phi) + b$, where $v$, $b$ and $\phi$ are fit parameters. The frequency varies with the detuning because of the varying overlap of the wavefunctions which controls the strength of $J$. The $T_2^*$ dependence on detuning is consistent with charge noise, as expected in this kind of qubit. The visibility $v$ is 65%, limited predominantly by state preparation and measurement errors. The quality factor $Q = fT_2^*$ is a measure of the speed/error rate ratio of logic gates and varies from 2 to 12 across the detuning range ($Q$ grows with detuning).

Figure 7: **(a)** Pulse sequence for AC-control. The qubit is operated in the $\{|\uparrow\downarrow\rangle, |\downarrow\uparrow\rangle\}$ basis. The detuning is modulated at the qubit frequency $f_{ac} \approx \Delta E_Z/h$ and the ac phase controls the rotation axis along the plane perpendicular to $\Delta E_Z$. **(b)** Rabi oscillations as a function of the excitation frequency $f_{ac}$ and manipulation time $t_{manip}$. **(c)** Rabi oscillations on resonance at $f_{ac} = 0.9$ MHz with a frequency $f_{Rabi} = 0.17$ MHz. The visibility is 65%, limited mostly by state preparation and readout errors, and is similar to that of exchange rotations. The decay is exponential with $T_2^* = 19$ μs. The quality factor is 3.2. Many optimizations are possible, including faster spin-orbit/Rabi drive and better preparation/readout. Results on dynamical decoupling will be presented in future work.

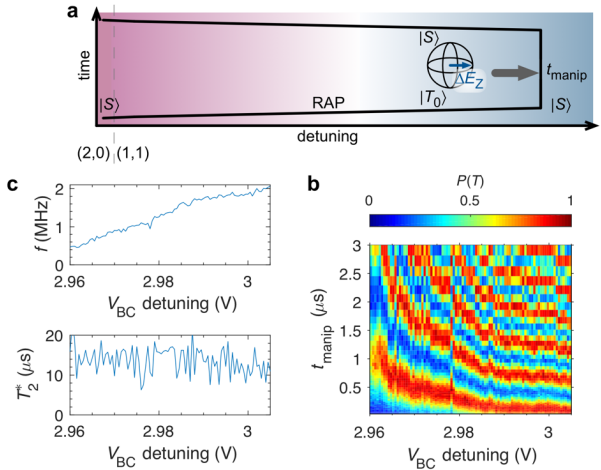

Figure 6: **(a)** Pulse sequence for DC-control of $\Delta E_Z$ rotations in (b). **(b)** Coherent spin-orbit-driven rotations in $P(T)$ versus the $V_{BC}$ coordinate of the detuning pulse and the manipulation time $t_{manip}$. **(c)** Frequency $f \sim \Delta E_Z/h$ and coherence time $T_2^*$ of the rotations extracted from fits to the data in (b) of the form $P(T) = \frac{v}{2}\exp\left(-(t_{manip}/T_2^*)^2\right)\sin(2\pi f + \phi) + b$, where $v$, $b$ and $\phi$ are fit parameters. The cause for the detuning dependence of the spin-orbit-driven rotations is not well understood. The $T_2^*$ is independent of detuning, as expected for noise from the residual $^{29}$Si magnetic fluctuations. Each vertical line is acquired over 2.8 minutes to determine $T_2^*$. Averaging over longer times saturates $T_2^*$ at $3.4 \pm 0.3$ μs after 2.2 h. The visibility $v$ is 75%, limited mostly by state preparation and measurement errors. The quality factor $Q = fT_2^*$ varies from 6 to 22 across the detuning range ($Q$ grows with detuning).